\documentclass{aa}
\usepackage{graphicx}
\usepackage{natbib}
\begin{document}
   \title{A general catalogue of 6.7-GHz methanol
     masers\thanks{The complete catalogue (Table \ref{tab:catalog}) is
     available 
     in electronic form at the CDS via anonymous ftp to
     cdsarc.u-strasbg.fr (130.79.125.5) or via
     http://cdsweb.u-strasbg.fr/cgi-bin/qcat?/J/A+A/ }}  

     \subtitle{I: data}
   \author{M. R. Pestalozzi 
          \inst{1}
          \and
          V. Minier \inst{2}
          \and
          R. S. Booth \inst{1} 
          }

   \offprints{Michele Pestalozzi \\
email: michele@oso.chalmers.se \\}

   \institute{Onsala Space Observatory, 439 92 Onsala, Sweden \\ 
              \email{michele,roy@oso.chalmers.se} 
                      \and
              Service d'Astrophysique, DAPNIA/DSM/CEA CE de Saclay,
              91 191 Gif-sur-Yvette, France \\ 
              \email{Vincent.Minier@cea.fr}            }
   \date{Received 12 December 2003 / Accepted 10 November 2004}

   \abstract{Methanol masers are often detected in regions of intense star
     formation. Several studies in the last decade indicate that 
     they may even be the earliest signpost of a high-mass star-forming
     region. Their powerful emission make them very good candidates for
     observations using both single--dish telescopes and 
     interferometers, the latter allows detailed structural and dynamical
     studies of these objects. We have prepared 
     a catalogue of all known 6.7-GHz methanol masers, discovered both
     by surveys that targeted possible associated objects and unbiased
     surveys covering a large 
     fraction of galactic longitudes across the Galactic plane ($-0.5\degr
     \le b \le 0.5\degr$ for most of the regions). The catalogue contains 519
     sources which are listed with their kinematic 
     (galactocentric and heliocentric) distances as well as possibly associated
     IR objects. We find that 6.7-GHz methanol masers clearly trace the
     molecular ring of our Galaxy, where most of the OB associations are
     located. The
     present list of masers also reports detections of other masing
     transitions of methanol as further information for the study
     of the maser phenomenon. In a further publication we will address some
     statistical considerations.

   }

   \maketitle
%
%________________________________________________________________

\section{Introduction}

Methanol masers were originally detected toward regions of active star 
formation (e.g. GMCs, \ion{H}{ii} regions) and have recently gained increasing
interest in the community. The first observed transition was the
$J_2{\rightarrow}J_1$-E line at 25~GHz, which was discovered toward Orion-KL by
\citet{bar71}. Nearly 
20 years later, \citet{bat87} and  \citet{men91a} detected two
very bright (1-1000 Jy) methanol maser transitions at 12.2 and 6.7~GHz,
respectively. 
Given that methanol masers at 6.7 and 12.2~GHz were associated and had never
been observed toward 25-GHz maser sources, \citet{men91b} proposed to
empirically 
divide methanol masers in two classes, {\it Class I} (e.g. 25 and 44 GHz
masers)  
and  {\it Class II} (e.g. 6.7 and 12.2 GHz). This classification is still
valid. Class II masers  
are directly associated with the birth place of a massive star (e.g. hot
cores, ultra-compact  
\ion{H}{ii} regions) and may be pumped radiatively by infrared radiation from
nearby warm dust  
(\citealt{sob97}). Class I masers are seen offset from strong radio continuum
sources (i.e. young ionising high-mass stars) and may be pumped collisionally 
(\citealt{cra92}). However, the nature of Class I maser sites remains poorly
known. 

During the past decade, extensive maser
searches have been undertaken using two different 
strategies: (1) targeted searches undertaken toward colour--selected
infrared (IR) sources and known regions of intense star formation (e.g. IRAS
sources, OH and H$_2$O masers); (2)
full-sampling and unbiased surveys covering large regions of sky conducted
both in the Northern and in the 
Southern hemispheres. These methods yield different results in terms of
detection statistics. The targeted surveys focused only on well-known sites
and missed masers arising in regions where they were not necessarily expected. 
The unbiased searches detected all
sources in a certain area of sky, and were biased only by the sensitivity of
the observations. 

To date, the most powerful and widespread methanol maser transition is the
$5_1{\rightarrow}6_0$-A$^+$ 
line at 6.7 GHz. It plays a key role in finding new high-mass star-forming
regions and in studying their kinematics at high angular resolution.  
The objective of this paper is to provide a complete
catalogue of the 6.7-GHz methanol maser sources. Section \ref{sec:mainres}
describes the searches 
for methanol masers and summarises their main results. Section \ref{sec:cat} 
describes the catalogue and lists the referenced publications.
Section \ref{sec:concl} gathers the conclusions we draw after
the first analysis of the data in the catalogue.
Further statistical analysis of the catalogue will be the
matter of a further publication. 

\section{Searches for 6.7-GHz methanol masers: 1991-2004}
\label{sec:mainres}

\subsection{Targeted surveys}

6.7-GHz methanol masers were discovered by \citet{men91a}. At this frequency
the author expected to find an enhanced absorption toward Class I maser
sources. The observation of a sample of 123 star-forming regions including 
both Class I and Class II maser regions as well as \ion{H}{ii} regions and OH
masers yielded instead 80 detections of a strong maser emission, which was
characterised as a Class II maser. All previously known Class II methanol
maser sites in that sample showed 6.7-GHz maser emission, as did 78 of the
98 main line OH maser sources in the same sample.  

The work by \citet{men91a} was a starting point for a number of extended
targeted searches for 6.7-GHz methanol masers, which probed large samples of
putative star-forming regions. These targets were selected either by the
presence of 
maser species (OH, H$_2$O and 12.2-GHz methanol masers) or by IRAS point
sources exhibiting IR colours of ultra-compact (UC) \ion{H}{ii} regions
(e.g. \citealt{woo89}). The surveys were conducted by
\citet{mcl92a,mcl92b,gay93,schu93,cas95a,vdw95,vdw96,wal97,sly99,szy00a}. The
upper part of Table~\ref{tab:comp} lists these targeted searches with
their detection rate, the $1-\sigma$ sensitivity of the observations (when
reported in the associated publication) and the target types.  A list of all
searches at 6.7\,GHz is presented in Table \ref{tab:refs}.

\citet{mcl92a} reported a 100\% detection rate of 6.7-GHz methanol masers
toward known 12.2-GHz masers. All of the new sources were associated with
compact IR sources from the IRAS point source catalogue (IRAS PSC) within the 
positional errors ($\sim1$~arcmin). Subsequent work by the same
authors (\citealt{mcl92b,gay93}) found 40 and 18 new methanol masers
respectively toward
OH maser sources. \citet{cas95a} listed 245 6.7-GHz methanol maser sources
detected toward star formation regions, and OH maser sites with a detection
rate of $85\%$.  
From the luminosity function of the masers, the authors concluded that there
should be about 500 sources in the Galaxy.

Furthermore, methanol masers seemed to be associated with OH maser sources
having  the {\it reddest} far-IR colours ($\log_{10}(S_{100}/S_{60}) > 0.5$). 
\citet{schu93} searched for 6.7-GHz methanol masers toward colour selected
IRAS sources. The selection criteria were based on the \citet{woo89} IRAS
colour criteria for UC \ion{H}{ii} regions, with the addition of lower limits
for 60- and 100-$\mu$m fluxes. This survey suggested for the first time that
massive star formation regions could be traced by methanol masers
alone. Larger samples of colour-selected IRAS sources were searched by
\citet{vdw95}, where the properties of the 31 new sources as well as all the
previous detections are discussed. No correlation between maser and IRAS flux
densities could be found. Spatially, methanol masers seemed to be concentrated
in the inner part of the Galactic plane and followed the velocity curve of
the Milky Way. \citet{vdw96} noticed that methanol masers were concentrated in
the Galactic plane, showing a scale-height significantly smaller than the
molecular gas in the Galaxy. The luminosity function of methanol masers seemed 
to be a power law with an index close to --2. \citet{wal97} found that their
201 detections toward 535 IRAS 
colour selected sources lay between spiral arms. This result suffered from an 
ambiguity in the kinematics distances. Two extensive targeted searches
(\citealt{sly99}; \citealt{szy00a}) were undertaken
toward very large samples of IRAS sources up to Northern hemisphere
declinations. The detection rates were $\sim10\%$. \citet{sly99} pointed out
the large velocity dispersion of the sources and their uniform distribution
over the inner part of the Galaxy. \citet{szy00a} and \citet{szy00b} found
sources with IR colours falling outside the range for typical UC \ion{H}{ii}
regions. 

Surveys giving negative results have also been reported. 6.7\,GHz methanol
masers were not detected towards a sample of low-mass star-forming regions
(\citealt{min03}), of evolved stars (\citealt{koo88}) and starburst galaxies
(\citealt{phi98b}), although the line was detected in the Large Magellanic
Cloud (\citealt{elli94b}). Long term monitoring of 54 
methanol maser sources showed that most masers were variable over a few days
to several days (\citealt{goe04a}). The types of behaviour include
non-variability to periodic maser flares (\citealt{goe04b}).  

\subsection{Full-sampling unbiased surveys}

Parallel to targeted searches, a series of unbiased surveys covering large
areas of the sky were undertaken (Fig. 1, online). The main goal of these
observations  was to find all detectable sources in the surveyed
regions. These surveys were carried out by \citet{elli96}, \citet{cas96a},
\citet{cas96b}, \citet{szy02}, the group in Noto (Italy) (Palagi et al.) and
the group at the Onsala Space Observatory (OSO) in Sweden (\citealt{min00b},
\citealt{pes02a}). Table \ref{tab:blind} lists all galactic regions that have
been covered by unbiased observations.  

The OSO survey covered a total of about 55 square degrees of sky. The
portion of the Galactic plane usefully observable from Onsala extended to
galactic longitudes $l>40^{\rm o}$, in contrast to the other unbiased surveys,
which had focused on the inner parts ($320<l<40^{\rm o}$). This is illustrated
by the colour boxes in Figure 1. The OSO survey yielded 11
detections, 4 of them that are new. Only one detection was made for longitudes
greater than $80^{\rm o}$. 

The unbiased surveys have been very effective in detecting methanol maser
sources where no clear signposts of star formation were previously known. 
\citet{elli96} found that about 50\% of the 100 detections in his survey do
not have any clear IRAS PS counterpart. The same conclusion 
was confirmed by \citet{szy02}. Three of the four new sources of the
Onsala survey are of this kind as well.

\begin{figure}[!ht]
\begin{center}
\includegraphics[width=7.2cm]{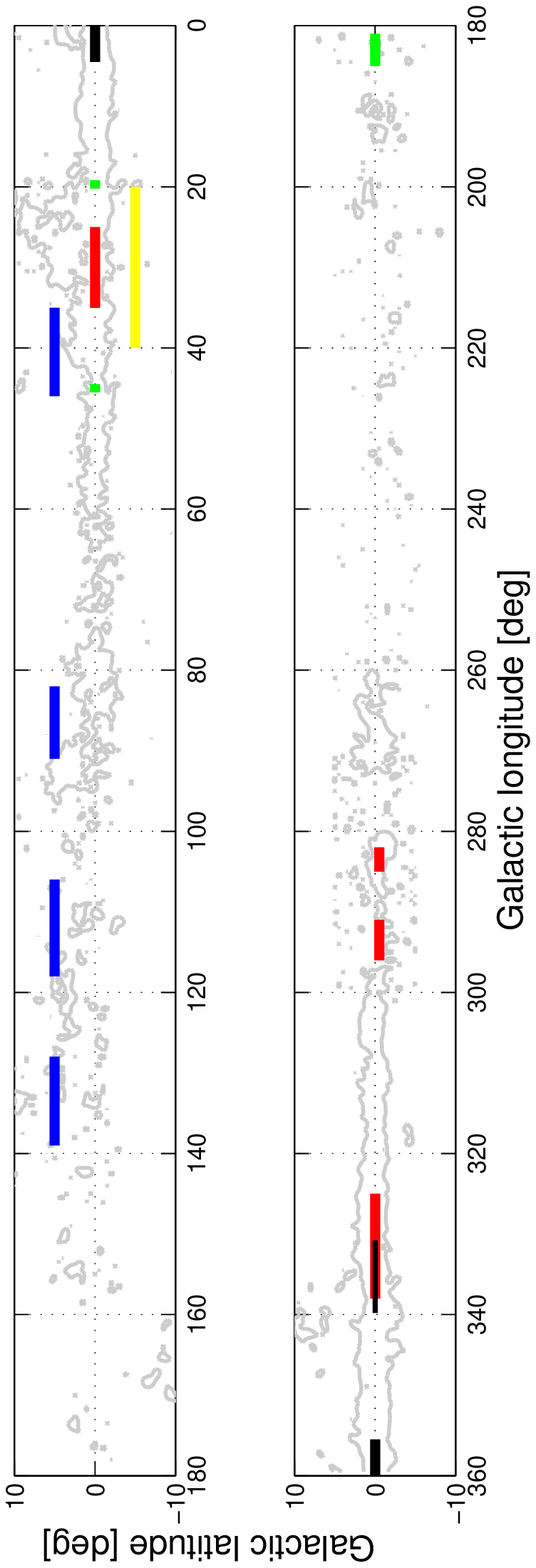}
\caption{Available online only. Map of the regions of the Galactic plane which
  were covered by 
  full-sampling observations. The grey contours are the 1\% CO emission level
  from \citet{dam87}. Blue are the regions observed by the Onsala group
  (offset at galactic latitude +5$^{\circ}$), red
  by \citet{elli96}, black by \citet{cas96a} and \citet{cas96b}, yellow by
  \citet{szy00a}, green by the group using the Noto telescope.}
\end{center}
\label{fig:blind_gal}
\end{figure}

\begin{figure}[!ht]
\begin{center}
\includegraphics[width=7.1cm]{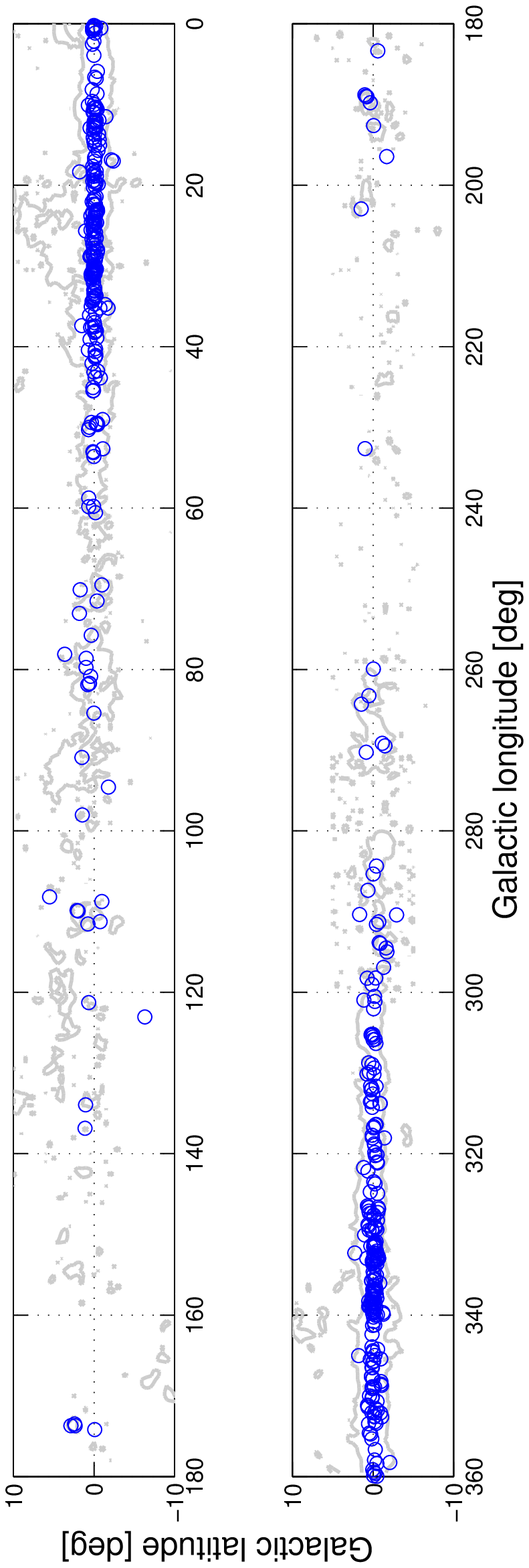}
\caption{Available online only. Detected 6.7-GHz methanol masers in the Milky
  Way. The gray contours 
are the 1\% CO emission from \citet{dam87}. Having limited the range of
galactic latitudes between -10$^{\circ}$ and +10$^{\circ}$ two sources are
missing. These are MonR2 (see e.g. \citealt{men91a}) and NGC~2024
FIR4 (\citealt{min03}). }
\end{center}
\label{fig:gal_sources}
\end{figure}

\subsection{Follow-up observations}

High angular resolution observations using the Australia Telescope Compact
Array (ATCA) and Very Long Baseline Interferometry techniques (VLBI) revealed
that the physical association of methanol masers with UC \ion{H}{ii}
regions does not hold for all sources. In general, as seen by 
\citet{wal98} and \citet{min01}, the methanol maser appeared to be clearly
offset from the main radio continuum peak marking the UC \ion{H}{ii}
region. The maser sites are therefore radio-quiet sources. \citet{min01}
proposed that methanol masers are mostly associated with hot molecular cores
and hyper-compact \ion{H}{ii} regions, which probably represent earlier stages
of high-mass star formation than the UC \ion{H}{ii} region phase. 

(Sub)Millimetre observations of methanol maser sites (\citealt{pes02b},
\citealt{wal03} and \citealt{min04}) demonstrated that methanol masers 
with no radio continuum emission were always detected in the mm continuum. 
The spectral energy distribution of these sources was consistent with that 
of an embedded, luminous, massive protostellar object (\citealt{min04}). 
IR observations were also performed (e.g. \citealt{wal99,goe02,deb00})
indicating that methanol masers were not always associated with the brightest
IR source but might arise from a nearby, deeply embedded source. 

Furthermore, methanol masers might trace Keplerian rotating
protostellar disks seen edge-on. The masers arise from an extended, flat
structure with a linear velocity gradient along it. The best example was
imaged and modelled in NGC7538-IRS1 (e.g. \citealt{min98,pes04b}). Other
objects also showed an ordered structure both in space and line-of-sight
velocity (\citealt{nor98,min01}). The cases for disks are not straightforward
in all sources and methanol masers could also be associated with outflows
(\citealt{min01}).  

Finally, \citet{min03} detected a weak 6.7-GHz methanol maser toward
NGC~2024, a region of intermediate-mass star formation in Orion B. The maser
was associated with a massive protostellar core that could host a nascent
high-mass star. Recent observations by \citet{vor04} have also revealed
methanol emission at 6.7~GHz in the OMC-1 region near the 25-GHz maser site,
i.e. a Class I maser source.  

\subsection{Methanol maser modelling}

Theoretical modelling was developed to produce synthetic emission spectra of
methanol masers at 6.7 and 12.2\,GHz. It was found that in order to reproduce
typical brightness temperatures  of $>10^{12}$\,K, an IR radiation
field is needed as well as ambient dust temperature of about 150\,K
(\citealt{sob97}; \citealt{cra01}, \citealt{sut01} for refinements). 
This radiation field plays a dual role in producing
methanol masers. It is required to release methanol from the icy mantles of
dust grains and to inject a sufficient amount of methanol into the gas phase,
reaching the needed column densities of $\sim 5\times
10^{17}$\,cm$^{-2}$. It also provides pump photons for the population
inversion.  These conditions are actually met in a region where young,
embedded massive ($>8$ M$_{\odot}$) stars are present (see \citealt{min03} and
references therein).  
The IR radiation field needed for pumping the masers could then be provided
either by warm dust near the protostar or by shocks produced by accreted
material and outflows. 

%__________________________________________________________________

\begin{figure}[!h]
  \begin{center}
    \includegraphics[width=8.5cm]{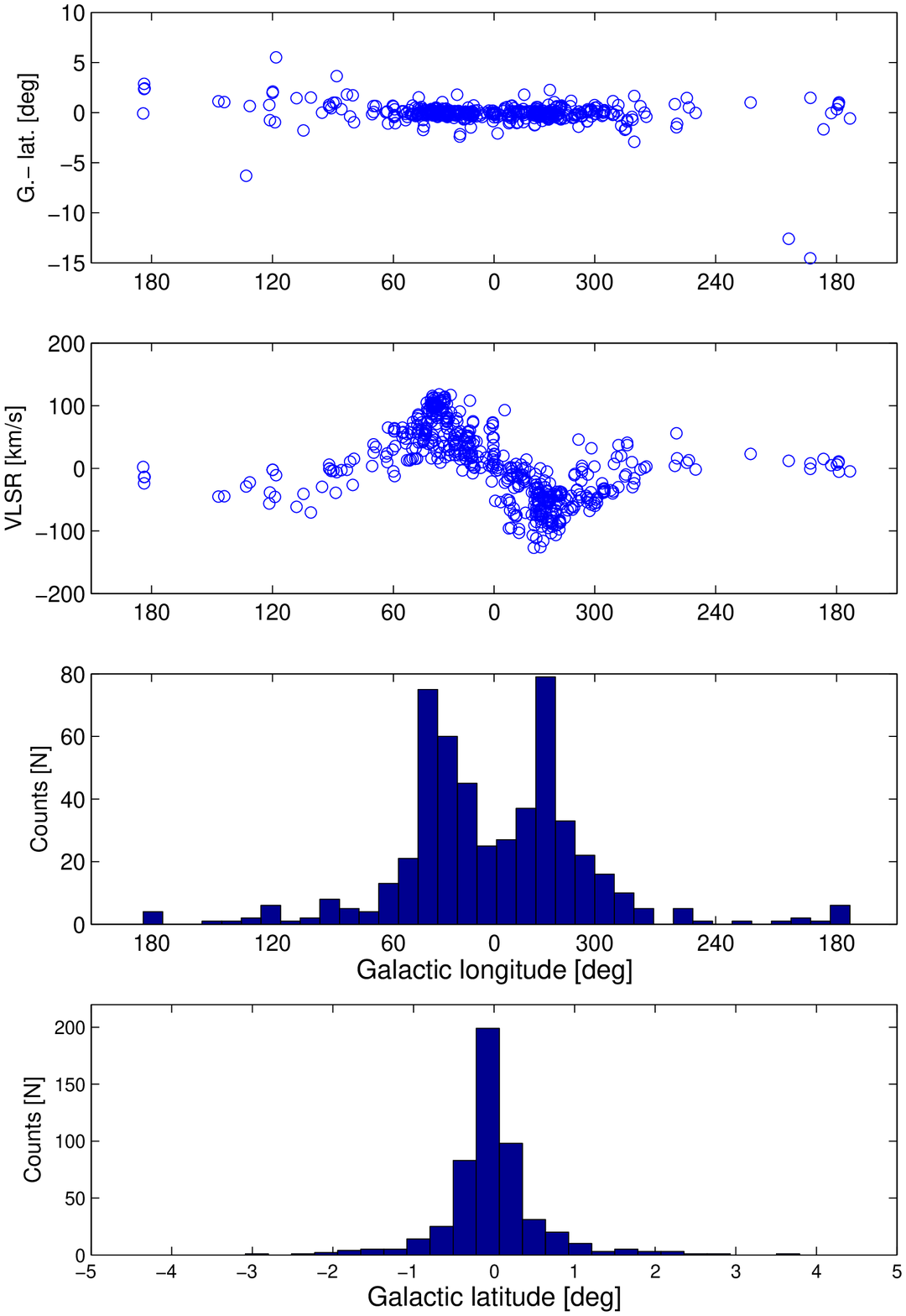}
    \caption{Distribution of the 6.7-GHz methanol masers in the Milky
      Way (top), distribution of the LOS velocities versus galactic longitude
      (middle) and source counts versus galactic longitude and galactic
      latitude (bottom). Notice the concentration of sources in a central ring
      of l=$\pm$50$^o$ and a stripe of about 1$^{\circ}$ across the Galactic
      Plane.} 
    \label{fig:n_gallon}
  \end{center}
\end{figure}

\begin{figure}[!h]
  \begin{center}
    \includegraphics[width=8cm]{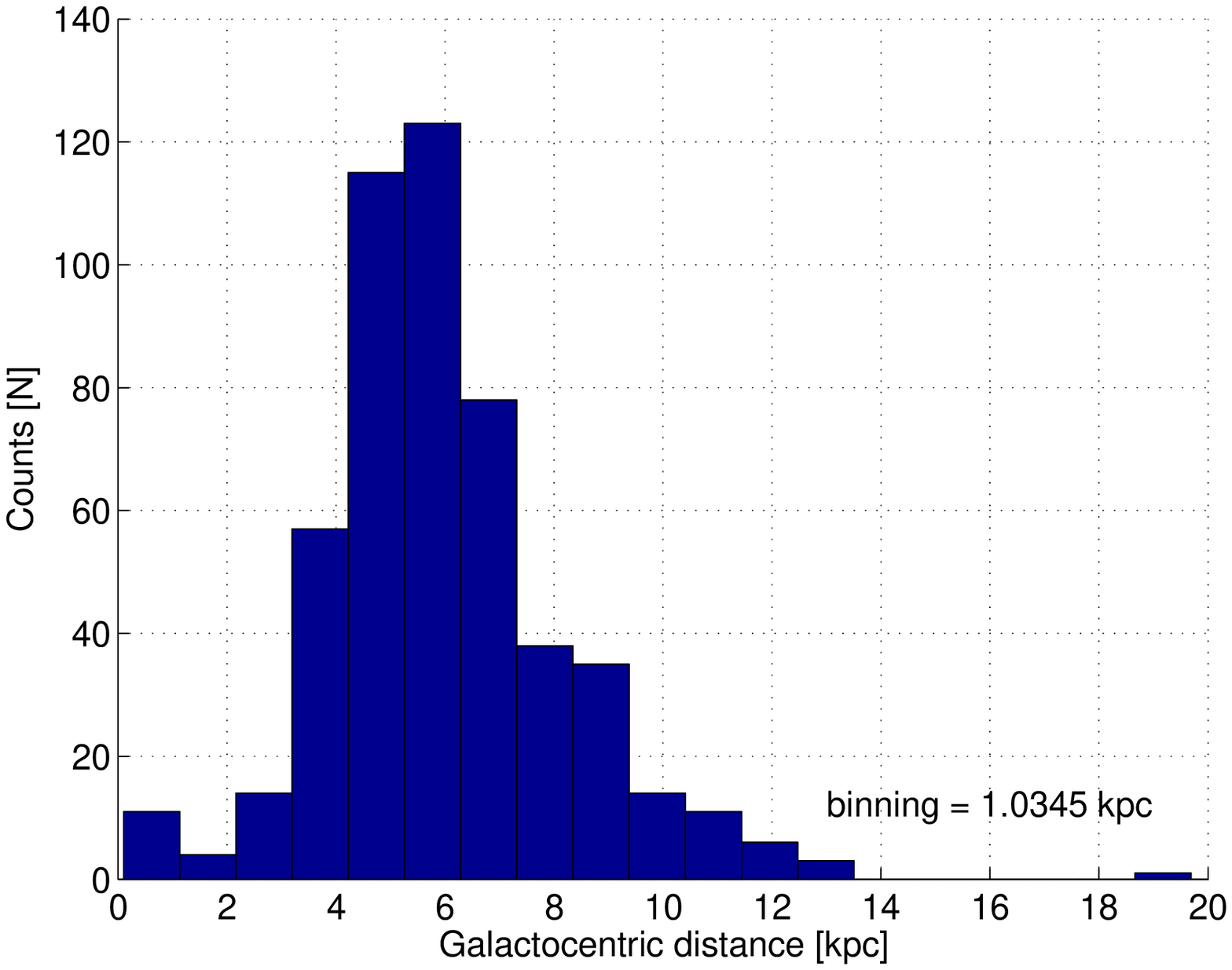}
    \caption{Distribution of the 6.7-GHz methanol masers versus
      galactocentric distance. Notice that the highest column is at the
      location of the molecular ring. } 
    \label{fig:n_gc}
  \end{center}
\end{figure}

\section{Catalogue description}
\label{sec:cat}

Each entry of the catalogue (Table \ref{tab:catalog}, available only in
electronic format at the CDS via anonymous ftp to
     cdsarc.u-strasbg.fr (130.79.125.5) or via
     http://cdsweb.u-strasbg.fr/cgi-bin/qcat?/J/A+A/) contains the
     following information: source name 
(\verb*|sounam|), right ascension and declination in J2000 coordinates
(\verb*|raj2000| and \verb*|decj2000|), galactic longitude and latitude
(\verb*|glon|, \verb*|glat|), radial velocity (\verb*|vlsr|), velocity range
with minimal and 
maximal velocities (\verb*|vmin|, \verb*|vmax|), peak flux (\verb*|peak|),
kinematic distances (galactocentric,  
\verb*|g_dist| as well as far and near heliocentric, \verb*|s_dfar|,
\verb*|s_dnear|), possible IRAS\footnote{Infrared Satellite,
  12,25,60,100\,$\mu$m, 2' pos. rms at 100 $\mu$m} and MSX\footnote{Midcourse
  experiment, 4, 8, 14, 21\,$\mu$m, 1-5" pos. rms} associated sources
(\verb*|iras|, \verb*|msx|), usual name (\verb*|u_name|), other masing
transitions detected (\verb*|m12,m44,m85,m86,m95,m107,m108,m156|), references
corresponding to the 
6.7-GHz methanol maser detections (\verb*|ref|). Celestial coordinates,
radial velocities and flux refer to the brightest spectral feature if the
source shows a complex spectrum.

The position accuracy may be deduced from the way the positions are listed in
the catalogue. Positions from single-dish measurements have one significant
decimal number in right ascension (RA) and none in declination (DEC). Australia
Telescope Compact Array (ATCA) positions from 
\citet{wal98} have been approximated to two significant decimal numbers in RA
and to the closest half second in DEC ($\pm$0.5"). Positions from
\citet{cas96a,cas96b} have been approximated to two and one significant decimal
numbers in RA and DEC, respectively. Full precision (three and two significant
decimal numbers in RA and DEC 
respectively) is listed for sources observed using VLBI
(\citealt{min00a,min01}). Not all source positions have been determined
accurately using high angular resolution, and this creates some difficulties
when trying to draw conclusions about associations. Note that many sources
show a number of features spread in space and frequency. When \verb*|wal98|
is listed in the references it means that the catalogue lists the position of
the brightest component in that region.

The flux densities listed in our catalogue refer to the first measurement if 
the sources have been detected several times. Since it is known that
many methanol maser sources show variability (e.g. \citealt{goe03,goe04a}), it
is difficult to be accurate to better than 10\%. 

The kinematic distances listed in the catalogue have been calculated using the
\cite{bra93} Galactic rotation curve, using $\rm R_0 = 8.5$\,kpc
(galactocentric 
distance of the Sun) and $v_0 = 220$\,kms$^{-1}$ (solar
orbital velocity around the Galactic center). However, we are not able to
safely discriminate between far and near heliocentric distance for each 
source. When available, it is possible to compare the listed heliocentric
distances with distances measured using e.g. optical or IR photometry. This is
not possible for all sources, since some of them are deeply embedded in a
thick dust cocoon and isolated from optical \ion{H}{ii}, so not visible 
at those wavelengths. What is more, some of the sources show {\it forbidden}
radial velocities, i.e. velocities not corresponding to the quadrant of the
LOS velocity -- longitude diagram they are in. For these sources the distance
is not calculated. Finally, sources having $90^{o} < l < 270^{o}$ show only one
heliocentric distance, as expected. The binning used in Figure \ref{fig:n_gc}
is 1\,kpc. This value was chosen after an error analysis of the calculation of
the galactocentric distance, which is affected by the uncertainties in the
galactic coordinates and the line--of--sight velocity (taken to be on average
1 arcmin in $l$ and $b$ and 0.2 kms$^{-1}$, respectively). 

For each source, possible associations with IRAS and MSX sources are given.
The IRAS associations often refer to the observed field where the IRAS point
source is taken as the name of the field. 457 sources are listed with an IRAS
counterpart. We list the closest MSX source in an area of radius
1, 2 or 3 arcmin from the position of the 6.7-GHz methanol maser. All but 7
sources show association with an MSX source. 370 sources seem to have an
MSX counterpart within 1\,arcmin, another 119 within 2\,arcmin. This fact comes
from the better angular resolution of MSX, which could prevent the problems
coming from confusion. Only one source (173.71+2.35) shows neither
an IRAS nor an MSX counterpart. Six sources have a clear IRAS, but no MSX 
counterpart, and 62 sources show an MSX counterpart without any IRAS
identification. In all but five of the last mentioned sources the MSX is
within 2 arcminutes.  

The references to the 6.7-GHz detections appearing in the catalogue are
listed in Table \ref{tab:refs} in order to facilitate the 
search. They always refer to 6.7-GHz detections or re-observations. 
When a source has been observed several times the reference to the listed
position is preceeded by a star.

The columns reserved for other masing transitions of methanol contain the value
of the peak flux density measured. The references to these detections are
listed in Table \ref{tab:other_masers}, but not in the catalogue. 

%_________________________________________________________________
% Table of the references and comparison

\begin{table}
\label{tab:catalog}
\end{table}

\begin{table}[h]
\caption{Summary of the detection rates of the searches used to compile the
  general catalogue together with the $1-\sigma$ sensitivity and the selection
  criteria used for the targeted surveys. 
  The information about the Noto survey is a private
  communication by Palagi et al.} 
  \begin{center}
    \begin{tabular}{llll}
\hline\hline
{\bf Targeted} & \verb*|#|det./\verb|#|obs. & 1$\sigma$ [Jy] &
Selection  \\ 
\hline
\verb*|men91| & 80 / 123 & -- & UC\ion{H}{ii}, H$_{2}$O, OH  \\
\verb*|mcl92a| & 19 / 19 & 4 & 12.2 GHz masers \\
\verb*|mcl92b| & 40 / 94 & 1.7-2 & OH, H$_{2}$O masers \\
\verb*|gay93| & 18 / 62 & 0.5 & OH masers \\
\verb*|schu93| & 35 / 235 & 1 & IRAS, col. sel. \\
\verb*|cas95a| & 245 / & 0.06 & OH (208), SFR \\
\verb*|vdw95| & 31 / 520 & 1.7 & IRAS (UC\ion{H}{ii}) \\
\verb*|vdw96| & 5 / 241 & 1.7 & IRAS (UC\ion{H}{ii}) \\
\verb*|wal97| & 201 / 535 & 1 & IRAS, col. sel. \\
\verb*|sly99| & 42 / 429 & 0.5-0.6 & IRAS (UC\ion{H}{ii}) \\
\verb*|szy00a| & 182 / 1399 & 0.7 & IRAS, col. sel. \\
%\verb*|cas95b| & 131 / 238 & & 12.2-GHz masers  \\
% & & \\
\hline
{\bf Unbiased} & \verb*|#|det./ (deg)$^2$ & & Detection rate \\
\hline
\verb*|cas96a,b| & 80 / 3.24 & 0.16 &  24.7 / (deg)$^2$
\\ 
\verb*|elli96| & 107 / 33 & 0.52 &  3.2 / (deg)$^2$ \\
\verb*|szy02| & 100 / 20 & 0.53  &  5 / (deg)$^{2}$
\\ 
\verb*|Noto00,02| & 10 / 4.5 &  -- &  2.2 / (deg)$^2$ \\
\verb*|Onsala99-03| & 11 / 50 & 1 &  $\sim$ 0.22 / (deg)$^2$  \\
\hline
    \end{tabular}
\label{tab:comp}    
  \end{center}
\end{table}

\begin{table}[h]
  \begin{center}
    \caption{References to the detection or high angular resolution
      observations of the 6.7-GHz methanol masers in the general
      catalogue. The reference codes refer to the same publications as in
      table \ref{tab:comp}.}  
    \begin{tabular}{llll}
\hline \hline
Code & Reference & Telescope \\
\hline
\verb*|men91| & \citealt{men91a} & Green Bank, 43\,m \\                  
\verb*|mcl92a| & \citealt{mcl92a} & Hartebeestoek, 25\,m \\               
\verb*|mcl92b| & \citealt{mcl92b} & Hartebeestoek, 25\,m \\               
\verb*|schu93| & \citealt{schu93} & Hartebeestoek, 25\,m \\              
\verb*|gay93| & \citealt{gay93} & Hartebeestoek, 25\,m \\
\verb*|cas95| & \citealt{cas95a} & Parkes, 76\,m \\                            
\verb*|vdW95| & \citealt{vdw95} & Parkes, 76\,m \\
\verb*|vdW96| & \citealt{vdw96} & Hartebeestoek, 25\,m  \\
\verb*|elli96| & \citealt{elli96b} & Parkes, 76\,m \\
\verb*|cas96a| & \citealt{cas96a} & ATCA \\
\verb*|cas96b| & \citealt{cas96b} & ATCA \\
\verb*|wal97| & \citealt{wal97} & ATCA \\
\verb*|wal98| & \citealt{wal98} & ATCA \\
\verb*|sly99| & \citealt{sly99} & Medicina, 32\,m \\
\verb*|min00| & \citealt{min00a} & VLBA, EVN, ATCA \\
\verb*|szy00| & \citealt{szy00b} & Torun, 32\,m \\
\verb*|szy02| & \citealt{szy02} & Torun, 32\,m \\
\verb*|min01| & \citealt{min01} & EVN \\
\verb*|min03| & \citealt{min03} & ATCA \\
\verb*|noto|  & Palagi et al. & Noto, 25\,m \\
\verb*|on00| & Onsala blind Survey & Onsala, 25\,m \\
\hline
    \end{tabular}
    \label{tab:refs}
  \end{center}
\end{table}

\begin{table}[h]
  \begin{center}
  \caption{Summary of the galactic regions covered by full-sampling
    observations searching for 6.7-GHz methanol maser emission. The first and
    last Noto-regions have been covered to 50\% and 75\%
    respectively. Some of the Onsala-regions have been expanded in
    area but not yet analysed. The expected detection rate in these regions is
    low and will not change our results significantly.}
  \begin{tabular}[h]{cccl}
\hline
\hline
 $l$-range & $b$-range & Area & Project \\
\hline
--4.5$^{\circ}$ -- 4.5$^{\circ}$ & $\mid$ b $\mid  \le 0.5^{\circ}$ &
1.8 & \verb*|cas96b| \\
19.2$^{\circ}$ -- 20.2$^{\circ}$ & $\mid$ b $\mid \le 0.5^{\circ}$ &
0.5 & \verb*|noto| \\
20$^{\circ}$ -- 40$^{\circ}$ & $\mid$ b $\mid \le 0.5^{\circ}$ &
20.0 & \verb*|szy02| \\
25$^{\circ}$ -- 35$^{\circ}$ & $\mid$ b $\mid \le 0.53^{\circ}$ &
10.0 & \verb*|elli96b| \\
35$^{\circ}$ -- 46$^{\circ}$ & $\mid$ b $\mid \le 0.5^{\circ}$ &
11.0 & \verb*|on00| \\
%46$^{\circ}$ -- 48$^{\circ}$ & $\mid$ b $\mid \le 0.5^{\circ}$ &
%2.0 & \verb*|on00|$\star$ \\
%40$^{\circ}$ -- 44$^{\circ}$ & 0.5$^{\circ} \le$ b $\le$ 1.0$^{\circ}$ & 2.0 &
%\verb*|on00|$\star$ \\
%40$^{\circ}$ -- 44$^{\circ}$ & ----0.5$^{\circ} \ge$ b $\ge$ ----1.0$^{\circ}$ &
%2.0 & \verb*|on00|$\star$ \\
44.5$^{\circ}$ -- 45.5$^{\circ}$ & $\mid$ b $\mid \le 0.5^{\circ}$ &
1.0 & \verb*|noto| \\
82$^{\circ}$ -- 91$^{\circ}$ & $\mid$ b $\mid \le 0.5^{\circ}$ & 9.0
& \verb*|on00|\\
%80$^{\circ}$ -- 82$^{\circ}$ & $\mid$ b $\mid \le 0.5^{\circ}$ & 2.0
%& \verb*|on00|$\star$\\
%80$^{\circ}$ -- 82$^{\circ}$ & 0.5$^{\circ} \le$ b $\le$ 2.5$^{\circ}$ & 4.0 &
%\verb*|on00|$\star$ \\
%102$^{\circ}$ -- 106$^{\circ}$ & $\mid$ b $\mid$ \le 0.5$^{\circ}$ &
%5.0 & \verb*|on00|$\star$ \\
106$^{\circ}$ -- 118$^{\circ}$ & $\mid$ b $\mid \le 0.5^{\circ}$ &
13.0 & \verb*|on00|\\
%118$^{\circ}$ -- 123$^{\circ}$ & $\mid$ b $\mid$ \le 0.5$^{\circ}$ &
%6.0 & \verb*|on00|$\star$ \\
%127$^{\circ}$ -- 128$^{\circ}$ & $\mid$ b $\mid$ \le 0.5$^{\circ}$ &
%1.0 & \verb*|on00|$\star$ \\
128$^{\circ}$ -- 139$^{\circ}$ & $\mid$ b $\mid \le 0.5^{\circ}$ &
12.0 & \verb*|on00|\\
%139$^{\circ}$ -- 150$^{\circ}$ & $\mid$ b $\mid$ \le 0.5$^{\circ}$ &
%11.0 & \verb*|on00|$\star$\\
%180.0$^{\circ}$ -- 181.0$^{\circ}$ & $\mid$ b $\mid$ \le 0.5$^{\circ}$
%& 1.0 & \verb*|on00|$\star$ \\
181.0$^{\circ}$ -- 185.0$^{\circ}$ & $\mid$ b $\mid \le 0.5^{\circ}$
& 3.0 & \verb*|noto| \\
%200.0$^{\circ}$ -- 201.0$^{\circ}$ & $\mid$ b $\mid$ \le 0.5$^{\circ}$
%& 1.0 & \verb*|on00|$\star$ \\
282$^{\circ}$ -- 286$^{\circ}$ & --1.03$^{\circ}$ $\le$ b $^{\circ}$
$\le$ 0.03$^{\circ}$ & 4.0 & \verb*|elli96b| \\
291$^{\circ}$ -- 296$^{\circ}$ & --1.03$^{\circ}$ $\le$ b $^{\circ}$
$\le$ 0.03$^{\circ}$ & 5.0 & \verb*|elli96b| \\
325$^{\circ}$ -- 338$^{\circ}$ & $\mid$ b $\mid \le 0.53^{\circ}$ &
13.0 & \verb*|elli96b| \\
330.8$^{\circ}$ -- 339.8$^{\circ}$ & $\mid$ b $\mid \le
0.08^{\circ}$ & 1.44 & \verb*|cas96a| \\
Other regions & & 4.0 & \verb*|on00| \\
\hline
  \end{tabular}
  \label{tab:blind}
\end{center}
\end{table}

\begin{table}[h]
  \begin{center}   
\caption{References to the detection of masing transitions of methanol at
      other frequencies than 6.7\,GHz.  }
    \begin{tabular}{ll}
\hline\hline
Reference & Telescope, freq. [GHz] \\
\hline
\citealt{bat87} & 12 \\
\citealt{koo88} & Hat Creek, 12 \\
\citealt{bac90} & Yebes-14m, 44 \\
\citealt{has90} & Haystack, 44 \\
\citealt{kal94} & Metsaehovi-13.7-m, 95 \\
\citealt{sly94} & Parkes, 44 \\
\citealt{val95} & OSO-20m, 107 \\
\citealt{sly95} & Kitt Peak, 156 \\
\citealt{cas95b}& Parkes, 12 \\
\citealt{val99} & Mopra, 107-108 \\
\citealt{val00} & Mopra, 95 \\
\citealt{cas00} & SEST, 107 and 156 \\
\citealt{min02b} & OSO-20m, 85,86,95,107,108,111 \\
\citealt{bla04} & Toru\'n, 12 \\
\citealt{elli03} & Mopra, 85.5 \& 86.6 \\
\hline
    \end{tabular}
    \label{tab:other_masers}
  \end{center}
\end{table}

%__________________________________________________________________

\section{Summary and conclusions}
\label{sec:concl}

As a part of the Onsala unbiased survey of the Galactic plane searching for
6.7-GHz methanol masers (\citealt{pes02a}), we have gathered all 6.7-GHz
methanol masers found 
since 1991 in a general catalogue. Our goal was both to 
deliver to the community a useful tool for selecting potential young high-mass
star-forming regions and to draw some statistical conclusions about the
maser phenomenon in our Galaxy. From the first order analysis of the general
catalogue, several conclusions can be drawn: 
\begin{itemize}
\item 519 methanol maser sites are present in the catalogue;
\item 6.7-GHz methanol masers show a non-random distribution in the Galactic
  plane, concentrated in a ring of $20 < |\,l\,| < 50^{o}$ (see Figures 2
  and \ref{fig:n_gallon});
\item a Gaussian fitting of the distribution across the Galactic plane shows
  a FWHM close to 0.5$^{\circ}$. This is less than found in the literature
  before 1996 (see Figure \ref{fig:n_gallon}, bottom). This
  could be due to the fact that unbiased searches (started in 1996) have
  concentrated on regions close to the Galactic plane;
\item the location of this ring seems to agree very well with the position of
  the molecular ring of the Milky Way as defined by e.g. \citet{bro00} as
  having a radius of $\approx 0.55 \,\rm R_{0}$ (see figure
  \ref{fig:n_gc}). The  
  ring is also the location of most of the OB associations: this fact supports
  the idea that methanol masers are closely associated with massive star
  formation;
\item about 50\% of the sources have flux densities between 10 and 200\,Jy,
  45\% of the sources are weaker than 10\,Jy, 27.9\% weaker than 5\,Jy. This
  fact suggests that the population of methanol masers in the Galaxy is
  probably dominated by weak emitters. An example comes from \citet{szy02},
  where the new sources discovered have a mean flux below 10\,Jy. 
\end{itemize}

Unbiased, high sensitivity searches seem then to be a useful way to complete
the general catalogue of galactic 6.7-GHz methanol masers.  
%__________________________________________________________________

\begin{acknowledgements}

We thank the technical staff at the Onsala Space Observatory for the excellent
support while surveying the Galactic plane. We also are very grateful to
Simon Ellingsen and Andrew Walsh for providing accurate positions for a large
number of sources. 

\end{acknowledgements}

\bibliography{methanol,varia,othermasers,starformation+IR,surveys_cat,tech,extra_gal} 
\bibliographystyle{aa}

\end{document}